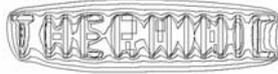

# FAILURE ANALYSIS AND FIELD FAILURES: A REAL SHORTCUT TO REALIABILITY IMPROVEMENTS

*G. Mura[a] and G. Cassanelli[b]*

[a] DIEE, University of Cagliari - INFM, Piazza d'Armi, 09123 Cagliari, ITALY
[b] DII, University of Modena and Reggio Emilia, Via Vignolese 905, 41100 Modena, ITALY

## ABSTRACT

Starting from two case histories, where only after thorough Failure Analysis the suddenly appearance of a failure was linked to much earlier events, the possibility of improving the reliability and of adjusting the reliability prediction tools are discussed.

## 1. INTRODUCTION

The role of Failure Analysis in the assessment of Reliability is widely acknowledged, and the most celebrated failure mechanisms have been granted along the years by the privilege of the introduction of process steps, in manufacturing microelectronics, that are solely dedicated to their prevention. Electromigration, corrosion, mobile charges, hot electrons, alpha particles, and latch-up are taught to microelectronic process engineers to explain the origin of maybe half of the current technological solutions.

Nevertheless, field failures still occur.

Handling field failures is in general a difficult task, that is even worst for those Small-Medium Enterprises (SMEs) that are small customers of the big microelectronic manufacturers: on one side, the overall small or medium volume itself of their production makes even few failures of statistical significance. Conversely, few field failures are not sufficient to adjust reliability figures, as failure rate, MTTF or π factors for any device in the most popular tools for Reliability prediction, whose direct application is of questionable utility just for those SMEs that are less armoured against reliability issues [1].

It remains the hope in miracles, that is in those highly accelerated methods that, often misunderstood, promise to increase the reliability of a system independent on its measurement. The problem is that field returns in general do not correspond to the failures produced during those tests.

The role of Failure Analysis will be here defended, in its deep acceptation, where a complete chain links the root failure mechanism to the observed failure mode (and not only a SEM picture is displayed to show some burnout). Detection of a specific root cause, when achieved, is that answer that is able both to address proper corrective actions in design and/or process and also to indicate more sound values for reliability parameters.

In this paper a couple of case histories in which the final damage of the component is associated to violent thermal phenomena are reported.

The two cases are linked by the common situation of having been mistaken in their diagnosis at the SME's plants, or even at some commercial Reliability Labs. In both cases sudden events have been reported, which was correct for the failure mode, although not necessarily for the failure mechanism. The butler of the situation was then claimed to be some transient Electrical Overstress (EOS) or some unexpected Electro-Static Discharge (ESD), and attention was paid to identify (and to remove) possible transients on currents, voltages, and electromagnetic fields during the operations of the two systems corresponding to the failure appearance.

The answers of the Failure Analysis were so subtly different to not completely destroy the first interpretation, but to back-date the occurrence of the root mechanisms well before the manifestation of the failure mode, and then displacing the focus from one class of operations of the failed system to some completely unsuspected others.

To avoid confusion on the focus of the paper, it is here anticipated that the well known Aluminium-Silicon interdiffusion mechanism will be called into play for both cases, and its main features will also be recalled.

Its role is only to witness the time delay between the first manifestation of the root mechanisms and the final failure modes.

## 2. CASE STUDIES

### 2.1 IGBT for racing car electronics

A popular commercial standard-speed IGBT, optimized for minimum saturation voltage and low operating frequencies, mounted in TO-220AB package, and employed in the on-board electronics of a racing car, failed after very few hours of intense operation.

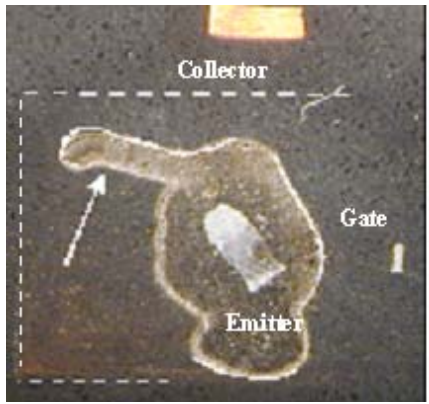

Fig. 1 Optical micrograph of the top surface after mechanical grinding. The dashed line defines the edges of the underlying chip. The view identifies the three sectioned metal connections (Emitter, Collector and Gate). The curvilinear lines delineate the carbonised plastic area, whose arrowed part belongs to the neighbourhoods of the original wire path.

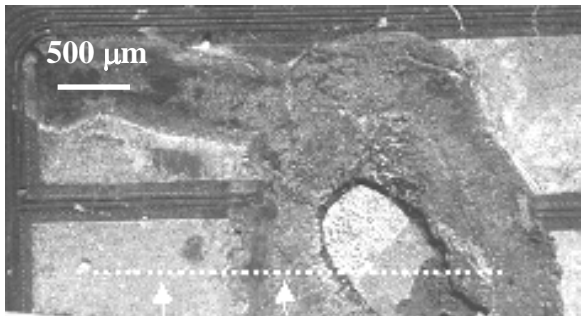

Fig. 2 DUT surface after the plastic removal. Pointed line indicates the direction of sectioning

More specifically, after one day of repeated tests on the race circuit and regular switching off at night, the system failed because of the short-circuiting of the IGBT as soon as it was switched-on, on the next morning.
The electrical measurements confirmed an overall short-circuit, with some tenths of an Ohm between Emitter (at the top of the chip) and Collector (at the substrate), and some 4 Ohms at each of them vs. Gate.

The first set of hypotheses pointed towards sudden stresses, like EOS or ESD due to some transient event at the switching-on. An accurate Failure Analysis was then started to find the evidence of the real mechanism. The first attempt to remove the plastic package by mixed grinding and chemical etching revealed (fig.1) an extended thermal damage of the plastic itself all around the thick Emitter wire. Carbonisation of the plastic mould prevented from uniform removal of the package, but a first strong indication for a catastrophic overheating was obtained.

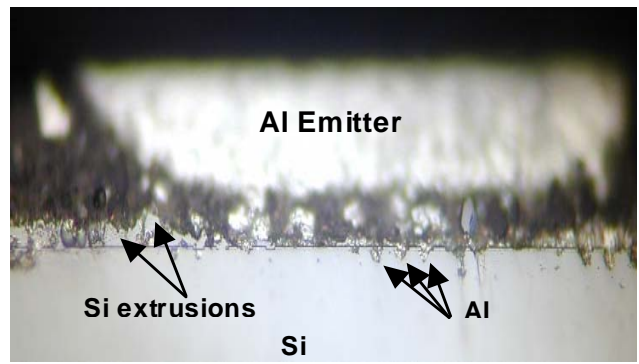

Fig. 3a General view of the sectioned area under the Emitter wire. Silicon extrusion into the Al metallization, and also Al intrusion into the Silicon crystal are displayed.

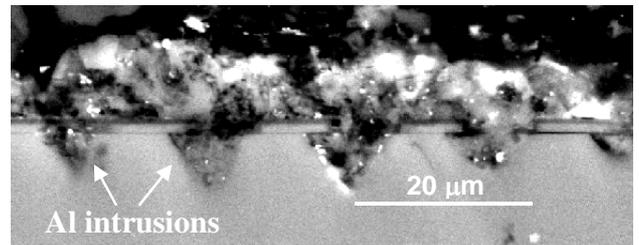

Fig. 3b SEM Detail of the Al intrusions. The polyhedric shape of the Al-filled cavities is a fingerprint of solid-state interdiffusion of Si into the overlaying metallic thin film.

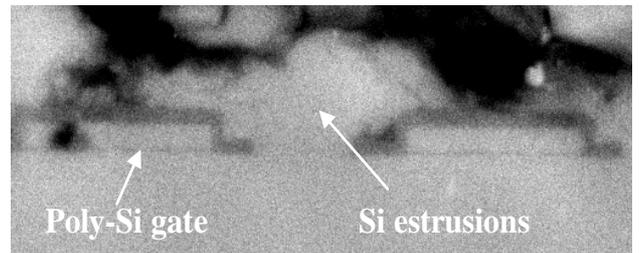

Fig. 3c SEM Details of the Si extrusions. The polysilicon gates (encapsulated in $SiO_2$) look intact.

The absence of any electrical connection on the top of the chip to the Collector clearly indicated the short circuit as a bulk failure. On the other hand, the x-ray EDS inspection of the exposed section of the wire revealed a central core nearly completely made of Silicon, surrounded by a region rich in Silicon and Oxygen.

The results were converging towards some strong vertical current flow, closer to some EOS event than to ESD, obviously the latch-up itself can be fired by a not destructive ESD event. Anyway, the dilemma of the sudden or progressive mechanism remains not yet resolved.

The complete removal of the plastic package was then carried out (fig.2), together with a cross-section of the damaged area (fig. 3). As a result, all the fingerprints

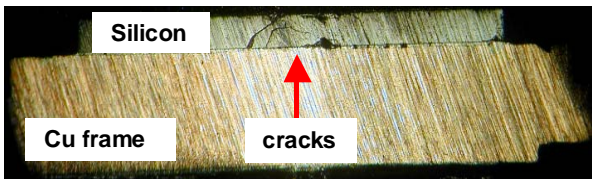

Fig. 4 section of the chip/thermal sinks structure. The creeks roughly crowd along the expected current/heat flux lines from the upper Emitter wire contact to the lower Collector surface.

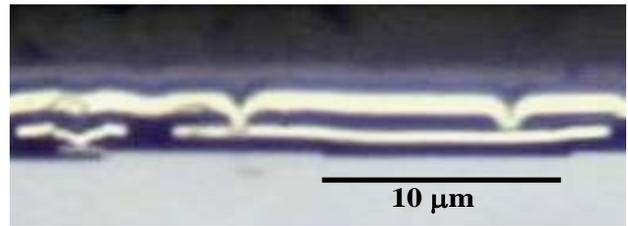

Fig. 5 Optical view of a cross sectioned device

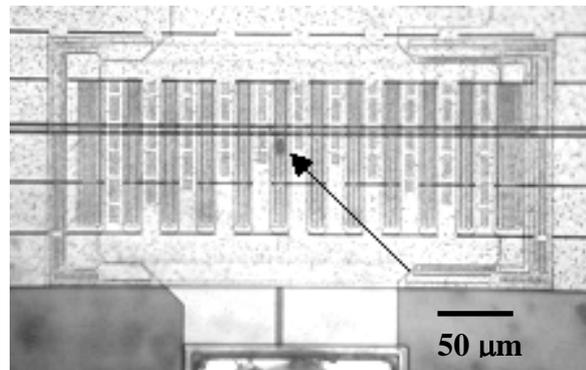

Fig. 6a General view of the surface of the failed input pin of the first chip, upon removal of the plastic package. The arrow points to a small "burnt" area.

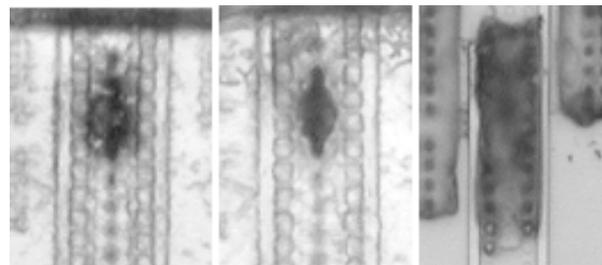

Fig. 6b Optical micrograph of the damaged area after sequential removal of polyamide (left) and nitride passivation (middle), and finally after HF (right)

of Al-Si interdiffusion were present and manifest (fig.3 a,b,c):
1) the reverse Al pyramids, hanging from those parts of the top Silicon surface, laterally limited by the SiO2 encapsulated polysilicon gates, where direct Al-Si contact makes the Emitter connection and
2) the Silicon precipitates inside the original upper Al thin film.

The key point is that the interdiffusion mechanism is not compatible with a sudden, transient phenomenon that would have resulted in fusion and chaotic re-crystallisation of the vertical structure. It should be argued that during the limited life of the device, an overcurrent was allowed to flow, undetected, able to expose the die to excessive temperature.

Another possibility is some missing control on thermal dissipation, mandatory for such a device. In that case the overdriving of the device could have raised the local temperature well beyond the specifications.

In any case, a strong and sustained current must be indicted for the failure, independent on the sudden, and late, manifestation of the final catastrophic step.

The origin of the overcurrent may be inferred by the statistics for IGBT devices, that indicate the latch-up mechanism as one the most frequent problems. The uncommon situation here is the relatively long survival of the device, up to its ultimate and irreversible short circuit, witnessed by the shown interaction at the wire contact and by the cracking of the die (fig.4).

### 2.2. CMOS Microcomputer

A single-chip 8-bit CMOS Microcomputer, employed on the control board of a boiler device for domestic application, failed in field operation, after passing the final production tests.

The failure analysis ran parallel to the analysis performed, on specimens with identical history, by the manufacturer, based on the results obtained by an external well-known reliability centre. Their conclusions were: "*It is considered that the destruction was caused by electrical overstress... check whether the supplied voltage at pin nn is stable*". In the manufacturer's report, the sentence was modified by deleting the focus on some possible voltage instabilities, and adding the specification "such as overcurrent" after the words "electrical overstress".

In both cases, the responsibility of the failure was completely loaded on the customer's shoulders and the corrective actions point only toward some not further specified controls on voltage or current.

A preliminary characterization of a device of the same lot revealed the structure of the multiple metal layer, where a very large upper metal layer (fig.5) run over the most part of the peripheral circuitry.

At a first glance, the optical inspection of the surface of one of the failed chips only showed minor damages (fig.6a), ill proportioned to the severity of the measured short-circuit: only a very small area showed some "burning" at the top metal layer in the region of the most

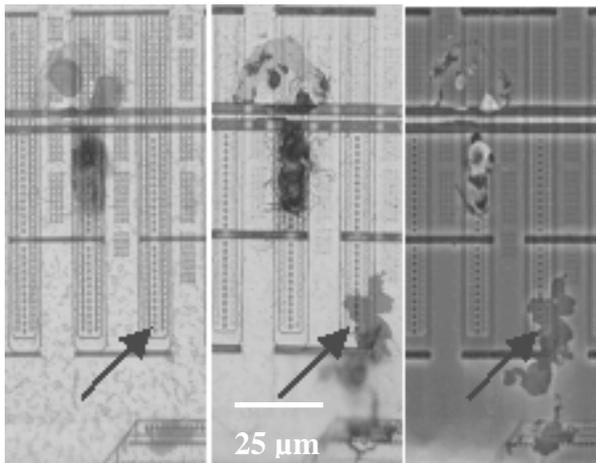

Fig. 7 Optical view of the damaged area after selective polyamide (a) and nitride (b) removal. The SEM image corresponding to b is displayed in (c). The arrows point the interdiffusion region on the top metal layer.

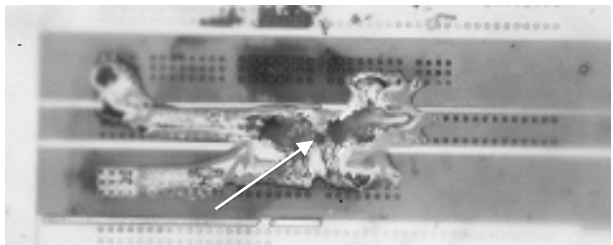

Fig. 8 (a) Optical view of the damaged area after complete HF etching.

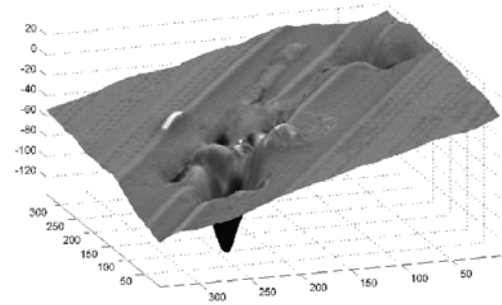

Fig 8(b) SEM 3D reconstruction [2] of the same detail, showing the central source of the Si material.

strongly failed input pin. No other evidences were found across the whole chip.

Some new information came during the chemical de-layering of the device (fig.6b): at the "burning" place, the nitride etch opened a hole into the upper metal layer, and the HF etch was not able to remove some islands of the lower metal layer. In general, this would be not strange at all, because some mix between silicon and aluminium, oxygen, possibly nitrogen and other elements is a normal effect of a thermal excess (fusion). The anomalous thing is that nothing looks melt: the upper metal layer displays a hole with indented boundaries, and the surviving "islands", rich of Silicon as expected, simply preserve, like a cast, the shape of the original aluminium, with its edges and reliefs.

The situation was even more evident on a second chip, where some more extended damage was observed at the surface of the failed pin. It is pointed out by the arrows in fig.7. The etch of the nitride passivation opened a hole (fig.7b) where the upper metal layer seemed intact (fig.7a). The edges of the hole strictly recall a polycrystalline structure, as expected for the aluminium layer. In fig 7b, and even more clearly in the SEM figure 7c, the hole shows an intact lower level metallization or, at least, a layer that preserves the original forms of that layer.

Upon HF cleaning of the surface (fig.8), the evidence arose for a central, small spot (arrowed) into Silicon from which a "Si-fountain" sprung from the substrate into the metal layers, branching into four different directions.
This ultimately allows, with the aid of fig.7 that compares the same region at different etching levels, to individuate the connection of the different structures of the damage: Silicon (amorphous) was present at the top metal level, which caused its removal under the etch of the passivation, while Aluminium penetrated the Silicon bulk, and was removed during the HF etch.

A 3D reconstruction [2] of the SEM images of the exposed Silicon surface (fig.8b) clearly shows the deep penetration of the damage (the interdiffusion "fountain"),

This is the other side of that same Al-Si interdiffusion phenomenon that appeared for the IGBT analysis. In this case, indeed, not the Al penetration into Silicon, but the progressive substitution of Al by Si atoms, coming from some deep contact point, to be found, is displayed. It is the same phenomenon whose seeds have been indicated for the previous IGBT case as "Si extrusions". The phenomenon is quite known, but it is always amazing to see the capability of Si to completely substitute Al, even preserving the shape of the metal layer.

It should be noticed as, once again, the evidence for Al-Si interdiffusion does NOT state that it is the root failure mechanism: the small spot clearly addresses towards ESD. Anyway, it is a secondary consequence, whose occurrence witnesses that the device operated a relevant time after the first failure firing before reaching the short-circuit state that flagged the failure.

Now, the interpretation of the failure is unavoidable: an original ESD event caused a local short-circuit, maybe not sufficient to directly bring the circuit to fail, but able to set up a small parasitic current, locally dense enough to fire the Al-Si interdiffusion mechanism.

The progressive evolution of this phenomenon then brought the device to its ultimate state at the observed time.

The relevant point is the casual occurrence of the discharge at some internal point of the circuits, but always under the wide upper metal layer. This suggested the possibility that not an I/O instability could cause the damage, but some capacitive coupling of the large parasitic capacitor, made of the upper metal layer and the substrate, and some huge electromagnetic disturbance near the device. The presence of a power switch very close to the board mounting the failed devices moved to introduce some specific protection (electrical shield and increased spacing between the critical elements), and the result was the complete disappearance of the problem.

## 3. DISCUSSION

The reliability prediction methodology proposed in [1] involves different approaches: empirical model prediction techniques, analysis of field return data about similar product, analysis of data about accelerated test and also *failure physics* of a single component, to asses the reliability of a new product.

The failure analysis of the CMOS microcomputer described in the second case study is used to improve the calculation of the corrective factor K relative to the control board in which the component is mounted. The information are used to decrease the value of $\lambda_{fs}$ that is the failure rate of the control board obtained through graphical and numerical analysis of the field return data.

The failure of CMOS microcomputer brings to the failure of the control board; but through the performed failure analysis this problem has been solved, so to asses the reliability of a new but similar board the $\lambda_{fs}$ has to be cleaned by this failure returns.

## 3. CONCLUSIONS

The Failure Analysis of devices in particular when failed in the field application, is not a deterministic science and enables us to establish only the most probable cause of the failure.

In both reported cases the sudden occurrence of the failure mode was not related to any sudden firing of the root failure mechanisms, but other hidden causes have been identified, with completely different corrective actions with respect to the first interpretations.

There is a simple and "correct" conclusion to this result: by means of thorough analyses, the first specimen (IGBT) was indicted for some higher sensitivity to latch-up, and the second (CMOS) to external-induced ESD events. This could move to correct the corresponding $\pi$ factors employed for calculating the actual failure $\lambda$, drawing a physically sound shortcut to the estimation of the reliability parameters for some critical devices of a given electronic system.

But there is also another conclusion, maybe less polite, that considers the documented case of the CMOS as a standard: when a SME faces a failure problem, the answer from the manufacturer will likely address to customer's responsibilities, possibly documented by some "official" report that displays a couple of pictures and indicts some "butler" as the culprit. Apart from any consideration about the integrity of intents of the parts, the problem exists of enabling the customer of a Failure Analysis to know if the paid job is good or not.

Ten years ago [3], a paper dealt with the problem of reliability of the Failure Analyses *[rue morgue]*, pointing out as many violations of a simple set of logical rules commonly occur also in authoritative Reports. A decade later, it seems that the situation did not improve, and that a whole class of customers remains undefended against a general lack of culture in reliability. A recall of those rules was worth of consideration in a recent paper [4].

**Acknowledgements**
The Authors wish to thank prof. F. Fantini and prof. M. Vanzi for the many useful discussions and for their continuous encouragement during the development of the reported study.